\documentclass[twocolumn,prl,aps]{revtex4}
\usepackage{epsfig,amsmath, amsthm, amssymb}
\def\beq{\begin{equation}}
\def\eeq{\end{equation}}
\def\be{\begin{equation}}
\def\ee{\end{equation}}
\def\ben{\begin{eqnarray}}
\def\een{\end{eqnarray}}
\def\beqa{\begin{eqnarray}}
\def\eeqa{\end{eqnarray}}
\def\eea{\end{array}}
\def\bea{\begin{array}}
\def\s{\,\,\,\,}
\def\Eo{E_o}
\def\m{m}
\def\m1{m_1}

\def\Er{E_{\cal R}}
\def\Es{E_{\cal S}}
\def\Sr{S_{\cal R}}
\def\Ss{S_{\cal S}}
\def\R{{\cal R}}
\def\S{{\cal S}}

\def\h1{h_1}
\def\h2{h_2}
\def\e1{e_1}
\def\e2{e_2}
\def\e{e}

\begin{document}

\title{Equilibrium temperature anisotropy and black-hole analogues}
 
\begin{abstract}
  When long-range interactions are present the
  usual definition of temperature implies that two systems in thermal
  equilibrium can be at different temperatures. This \textit{local} temperature
  has physical significance, if the sub-systems cease to interact, each system
  will be at their different local temperatures. This is formally related to 
redshifting of temperature in general relativity. We propose experiments to test this
  effect which are feasible using current microfabrication techniques. It
  is also possible to display thermodynamical analogues to black-hole
  space-time.
\end{abstract}

\author{Daniel K. L. Oi$^{(1)(2)}$, Jonathan Oppenheim$^{(1)}$}
\affiliation{$^{(1)}$Department of Applied Mathematics and Theoretical Physics, Wilberforce Road,
  University of Cambridge, Cambridge CB3 0WA, UK}
\affiliation{$^{(2)}$SUPA, Department of Physics, University of Strathclyde,Glasgow G4 0NG, UK}


\maketitle

\section{introduction}

When a system possesses long-range interactions, many familiar notions of
thermodynamics break down; the micro-canonical and canonical ensembles become
inequivalent~\cite{paddy}, there may be no stable equilibrium
configuration~\cite{antanov,lynden1}, and heat-capacities can be
negative~\cite{lynden1,thirring} as observed in fragmenting nuclei~\cite{negcap-nuc}
and atomic clusters~\cite{negcap-sod}. Formally, standard thermodynamics has
only been proved valid when interactions are short-range, or when the
long-range interactions are screened (e.g. plasmas)~\cite{landlieb}. There are
few methods for studying the thermodynamics of systems with long-range
interactions, although some models have been studied for special cases where
the thermodynamic limit exists~\cite{dyson69}, and
understanding them outside the standard framework using Tsallis or Renyi
entropy has been attempted.

Using a general formalism for studying such systems~\cite{ising} based
on techniques used in general relativity (where the equivalence principle is
exploited) -- an effect was noted which we summarise and extend as
follows. Consider many strongly interacting sub-systems in thermal
equilibrium. 
Using the standard definition of temperature (defined as the
\textit{local temperature}), each sub-system is at a different
temperature even though the entire system is at thermal equilibrium~\cite{ising}. Clearly
the standard definition does not satisfy a basic notion -- that it be constant
throughout the sample at equilibrium, yet it has a physical meaning -- if the
interaction is turned off suddenly and the sub-systems isolated, they will be
at their local temperatures. 
 The
observed temperature difference when a system is broken down into its parts is a property
of the system and is a function of its self-interaction.  What's more, the effect is 
formally related to effects found in curved space and black hole thermodynamics~\cite{ising}.
Here we show that this anisotropy of local temperature can
be observed, though this requires the long-range interaction to be strong
compared with the characteristic temperature of the sample. However, recent
advances in microfabrication may allow experimental access to thermodynamical
effects not found in macroscopic systems.  An experiment to measure the break-down of 
temperature in quantum systems was recently proposed in~\cite{hartmannmahler}. 

We review the basic formalism for analysing systems with long range
interactions and then apply this to a proposed experiment. We also show that one
can observe additional effects closely related to thermodynamical relations in
black-hole space-times.

\section{Long-Range Interactions}

Consider two sub-systems, with total energy $m$ of the form
\begin{equation}
m=E_1+E_2+G(E_1,E_2),
\label{eq:int}
\end{equation}
where $E_i$ are the local (non-interacting) energies and $G(E_1,E_2)$ is some
interaction potential (which may include self-interacting terms). As an
example, we consider the interaction of two clusters of classical spins in
local magnetic fields $b_1$ and $b_2$, with spin-spin interaction 
\begin{equation}
m=b_1 e_1 + b_2 e_2 -J_1 e_1^2/2 -J_2 e_2^2/2 -J_{12} e_1 e_2,
\label{eq:twoissings}
\end{equation}
where $e_1$ and $e_2$ are the spin-excesses of each cluster
$e_i=n_i^\uparrow-n_i^\downarrow$, ($n_i^\uparrow$ and $n_i^\downarrow$
being the number of up and down spins respectively of cluster $i=1,2$), $J_i$ are intra-cluster
couplings, and $J_{12}$ is the inter-cluster coupling. Such an
interaction arises from a standard interaction between spins of the form
\beq
m=\sum b_1 \sigma_1^j + \sum b_2 \sigma_2^k - 
\sum_{lm}\sum_{i,i'=1,2} J_{ii'}^{kl} \sigma_i^k \sigma_i'^l 
\eeq
with the spins of cluster $i$ being $\sigma_i^j$.  For small clusters, the 
interaction within each cluster can be
approximately the same for all spins, not just nearest neighbour, i.e. 
$J_{11}^{kl}=J_1, J_{22}^{kl}=J_2$, and the
interaction between each cluster is approximately uniform over each cluster
i.e. $J_{12}^{kl}=J_{12}$.  
We then drop constant terms to obtain Eq.~(\ref{eq:twoissings}).
Gravity is another common
example of such an interaction, where the potential term depends on the local
energies (masses) of each system.

One can define the temperature of one of the clusters
(the system $\S$), by making the other system very large (the reservoir $\R$).
If the total energy $m$ is fixed, in this limit the probability that the
system has energy $\Es$ is given by~\cite{ising}
\begin{equation}
p(\Es)d\Es=\sum_E \Omega_S(\Es)\Omega_R(E) e^{-\Es\beta_E}d\Es/Z_m
\label{eq:sume}
\end{equation}
where the sum is taken over all $E=\Er+\Es$ consistent with total energy
$m=E+G(\Es,\Er)$ and
\begin{equation}
Z_m=\int_m d\Es \Omega_S(\Es)e^{\Sr(\Er)}
\end{equation}
i.e. $Z_m$ is the total number of states at fixed total energy $m$.  We then
define the inverse temperature $\beta_E$ in the usual manner in terms of the
local extensive entropy
\begin{equation}
\beta_E\equiv\frac{\partial \Sr(\Er)}{\partial \Er} \s
\label{eq:localt}
\end{equation}
We shall refer to $\beta_E$ as the \textit{local temperature}.  The
motivation for using this term comes from general relativity.

Note that the temperature of the system is defined in terms of the derivative
of the \textit{reservoir's entropy}.  In the non-interacting case, no issues
arise from this definition: if two systems are in thermal contact in the
microcanonical ensemble, then $\frac{\partial \Ss(\Es)}{\partial \Es} \simeq
\frac{\partial \Sr(E)}{\partial E}$.  This is also true when the reservoir has no
long-range interactions, or when
the division of a \textit{single} system into a reservoir and
smaller system is purely formal (as we will see from symmetry considerations).
In general however, this is not necessarily true -- a
point which will be discussed shortly. One therefore should keep in mind that
the temperature is a property of a reservoir -- it gives the distribution
associated with a smaller system in contact with it.
Finally, note that the local temperature, as defined, is a function of $E$, we
make this explicit by writing $\beta_E$. There will be different
``temperatures'' depending on the value of $E$ the entire system is found in.

Let us now show that at equilibrium, a system can have an anisotropy in local
temperature.  Here we consider a slightly different situation and derivation than in
\cite{ising}. For concreteness, one could consider a system, as above, composed of two
clusters as above labelled $1$ and $2$. For fixed $E_1$ and $E_2$, the spin
excesses of each cluster are fixed, thus their respective entropy is also
fixed and given by
\begin{equation}
S_i(E_i)= - n_i^\uparrow\log\frac{n_i^\uparrow}{n_i^\uparrow+n_i^\downarrow}
- n_i^\downarrow\log\frac{n_i^\downarrow}{n_i^\uparrow+n_i^\downarrow}
\label{eq:spinentropy}
\end{equation}
and since each entropy is determined only by the spin excess, the total
entropy is additive, i.e. the number of states accessible to two systems is given
by the number of states accessible to one system, times the number of states accessible
to the other.  I.e.
\begin{equation}
\Omega_{12}(E_1,E_2)=\Omega_1(E_1)\Omega_2(E_2)
\end{equation}
where $\Omega_{12}$ is the total number of states of the combined system, and $\Omega_i$ is
the number of states accessible to system $i$ (the entropies of each system are just 
$S_i(E_i)=\log \Omega_i (E_i)$).
Additivity of entropy holds because we will find that we only need it at each value of $E_1,E_2$ 
and in this case, the entropy is additive 
in many systems because often the only correlations in a system
are related to correlations in energy.  We can thus consider more general systems with additive
entropies (when conditioned on local energy).  We now consider these two systems in contact with
a reservoir $\R$ at fixed total energy $m=E_1+E_2 + E_\R + G(E_1,E_2)$ and again taking the total local 
energy as $E=E_1+E_2+E_R$.  The probability that the two systems have energy $E_1$ and $E_2$ for given 
local energy $E$ is $\Omega_1(E_1)\Omega_2(E_2)\Omega_R(E_R)/Z_E$, and the probability that the local energy
is $E$ is denoted by $p(E)=Z_E/Z_m$, with the partition functions 
$Z_m=\sum_{E_1,E_2,E} \Omega_1(E_1)\Omega_2(E_2)\Omega_R(E_R)$ at fixed $m$ and 
$Z_E=\sum_{E_1,E_2}\Omega_1(E_1)\Omega_2(E_2)\Omega_R(E_R)$ at fixed $E$.
Then the probability that   
system $1$ has energy $E_1$ is
\begin{align}
p_1(E_1)
&= \sum_{E,E_2} p(E_1,E_2|E)Z_E/Z_m\notag\\
&=\sum_{E,E_2} \Omega_1(E_1)\Omega_2(E_2)\Omega_R(E_R)/Z_m\\
&\simeq   \frac{\Omega_1(E_1)}{Z_m}e^{S_R(m)-\beta_oE_1}
\sum_{E,E_2}
\Omega_2(E_2) 
e^{-\beta_o (E_2+G(E_1,E_2))}  \notag
\end{align}
where 
we have 
used the approximation that $E_\R \gg E_1+E_2+G(E_1,E_2)$ to expand $S_R$ around $m$ with the
inverse {\it global temperature} defined as 
$\beta_o\equiv\partial S_R(m)/\partial m$.  Since $m$ is held fixed, an absence of interaction $G(E_1,E_2)$
would give the usual fact that $p_1(E_1)$ will be
proportional to $\Omega_1(E_1)e^{- \beta_o E_1}$, and symmetrically for $p_2(E_2)$ -- thus the systems will be
at equal temperature.  Here,
we have the additional sum over $E$ and the factor due to the interaction term, which makes
the distribution of system $1$ and system $2$ asymmetric. 
Expanding $G(E_1,E_2)$ around $(E_1,E_2)=(\bar{E}_1,\bar{E}_2)$  the average energies, gives
\beq
p(E_1)\propto
\Omega_1(E_1)e^{-(1+\frac{\partial G(\bar{E}_1,\bar{E}_2)}{\partial \bar{E}_1}) \beta_o E_1}
\label{eq:twocan}
\eeq
which looks like a canonical distribution.  By symmetry, each system thus behaves
as if it has an
inverse temperature
\beq
\beta_i
=
(1 +   \frac{\partial G(\bar{E}_1,\bar{E}_2)}{\partial
      \bar{E}_i})\beta_o
\label{eq:tempint}
\eeq
One can calculate that this local temperature matches the standard definition of temperature given by 
Eq.~(\ref{eq:localt}) and we thus see that there is a 
temperature anisotropy whenever 
$\partial G(\bar{E}_1,\bar{E}_2) /\partial \bar{E}_2
\neq \partial G(\bar{E}_1,\bar{E}_2) /\partial \bar{E}_1$


Let us now understand the physical meaning of this local temperature.
Eq.~(\ref{eq:sume}) gives the probability distribution of a system $\S$ in terms
of the local temperature $\beta$, and the local energy $\Es$.  However, the local energy is not a
conserved quantity, and does not contain the interacting term.  To see what 
the locally conserved energy is,
we can expand $m$ to first order 
$m (E_1,E_2)\simeq m(\bar{E}_1,\bar{E}_2) +  \frac{\partial m}{\partial E_1}(E_1-\bar{E}_1) 
+  \frac{\partial m}{\partial E_2}(E_2-\bar{E}_2)$.  Since constant terms can be ignored, 
$ E_o^i\equiv \frac{\partial m}{\partial E_i}E_i$ can be identified with
the energy of each system in the presence of the mean
field due to the other interacting systems and is  
$(1 +   \frac{\partial G(\bar{E}_1,\bar{E}_2)}{\partial \bar{E}_i}) E_i$.  
In the case of a single system and reservoir, this energy is 
$\Eo\equiv \Es\frac{\partial m}{\partial \Es}$ and serves 
to tell us how we should define the locally conserved energy.
As is expected for a non-extensive system, the quantity is also non-additive,
but, as with the local temperature, will correspond with what happens in general relativity.

$\Eo$ is also a function of $E$ but we will not write this explicitly. 
%
With respect to energy levels $\Eo$, each system acts approximately as if it
is at inverse temperature $\beta_o$ (although 
in actual fact it is a mixture of temperatures~\cite{ising}).
%
%
We can think of $\Eo$ as the effective energy,  i.e. it is the energy of the
system in the presence of interaction with another system, and  $\beta_o$ can
be thought of as the closest thing one has to a physical temperature -- looking
at Eq.~(\ref{eq:tempint}) we see that the global inverse temperature
$\beta_o$ will be equal for both systems. Note however, that the global
temperature is not an intensive quantity.

If the interaction term is suddenly ``turned off'', (one can imagine
that the spins are suddenly separated so that they
no longer interact), then the energy that is ascribed to each energy level is no
longer $E_o^i$, but rather its local energy $E_1$ or $E_2$.  If the change
happens extremely quickly then the overall state of the system will not
change. Since
\begin{equation}
\beta_i E_i=\beta_oE_o^i
\label{eq:importantrelation}
\end{equation}
and the new energy of each system is now $E_i$, the measured temperatures will
be $\beta_i$ -- the local temperature.  In these systems, the local temperature
becomes the physical one when the interaction is turned off, which is analogous
to the fact that 
in general relativity, the local temperature
is the physical one measured by free-falling observers (for whom the gravitational 
interaction is ``turned off'').

Using the two coupled Ising models of Eq.~(\ref{eq:twoissings}) we would get a
temperature difference of \beq \beta_2 = \beta_1
\frac{1-J_{12}e_1/b_2-J_2e_2/b_2}{1-J_{12}e_2/b_1-J_1e_1/b_1} \s .
\label{eq:isingtempdif}
\eeq




Now if initially these two systems (or clusters), are far apart, and at equal
global temperature, one cannot push them together both adiabatically and
isothermally (constant global temperature) as is possible in the
non-interacting case. This can be seen from Eq.~(\ref{eq:importantrelation}).
Moving the systems together adiabatically requires keeping $\Eo\beta_o$ fixed,
but since $\Eo$ changes when $J_{12}$ becomes significant, one cannot keep
$\beta_o$ constant.  By recalculating $\Eo$ one can therefore calculate the
new global temperature. We see therefore that the global temperature is not an
intensive quantity.

\section{Black-hole on a benchtop}

To obtain an analogue of a black-hole,
we consider a single cluster of spins
with long-range self-interaction.  This can be obtained by setting $h_2=J_{2}=J_{12}=0$.
This gives a relation between the local and global temperature of
\beq
\beta
=\pm\beta_o\sqrt{1-\frac{2J_2 m}{h^2}} \s.
\label{eq:bhtemp}
\eeq
The positive solution looks exactly like the relationship between global and local temperature
in the Schwarzschild black-hole space-time due to redshifting, 
if we set the radius $r=h^2$ and
equate the spin-spin coupling with Newton's constant $J_1=G$.  Here $\beta_o$ 
is the temperature at infinity, while $\beta$ would be the local temperature measured
by an observer sitting close to the black-hole horizon.  In the analogue, the 
horizon is real -- for fixed $m$, there is no value of spin-excess 
which allows $r<2Gm$ i.e. $m>h^2/2J_1$.  There are also many similarities between this
analogue and a black hole in terms of the
way the entropy and energy behave, and we refer the reader to \cite{ising,area} for
details.
%

\section{Experimental Realisation}

Ideally, we would like to study these phenomena in complex macroscopic systems.
However these usually do not possess a strong enough long-range interaction to
produce an appreciable effect, hence it is necessary to go to small systems
where the the coupling between sub-systems can be comparable to the local
energies and low temperatures are easier to achieve.

\begin{figure}
\includegraphics[width=0.45\textwidth]{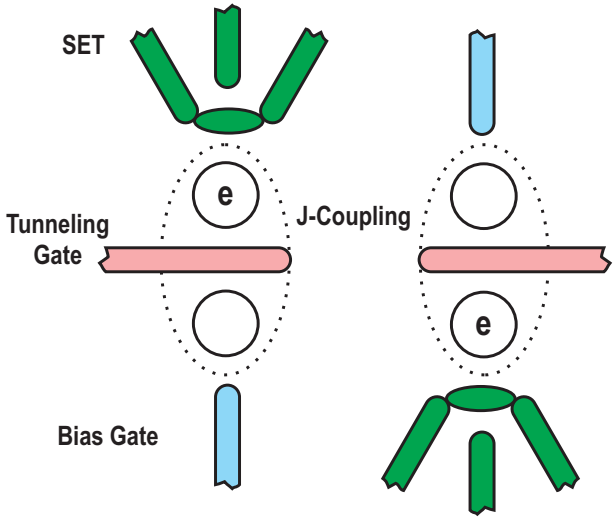}
\caption{Coupled double quantum dot (DQD) system constructed from donor
  phosphorus atoms in silicon with $20nm$ vertical and $30nm$ lateral
  separation. Each left and right DQD pair is initialised with one electron
  shared between the two dots.  The local Hamiltonians can be controlled by
  electrodes affecting the tunnelling rate and bias between the two dots. The
  interaction is of the Ising type ($J=2.9meV$) and single electron
  transistors (SETs) measure the localisation of the electrons to the upper or
  lower dots of each pair.}
\label{fig:DQDSys}
\end{figure}

\begin{figure}
\includegraphics[width=0.45\textwidth]{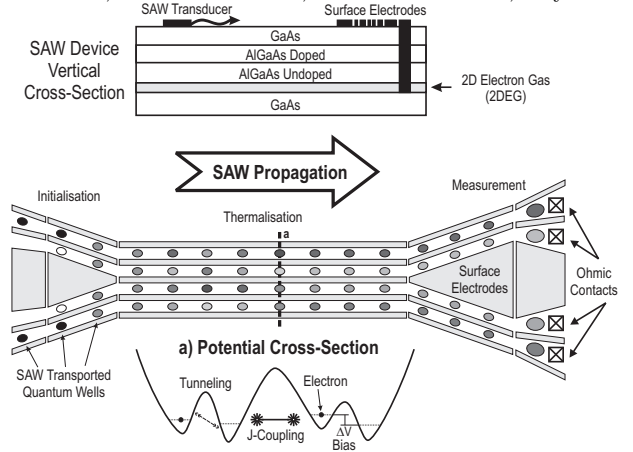}
\caption{Surface acoustic wave (SAW) device. The SAW created by an
  interdigital transducer produces a sinusoidal piezoelectric field in the
  direction of propagation which, together with surface electrode located 
  above a 2DEG, create a pattern of moving quantum wells which can be manipulated
  by static voltages.
  The advantage of this configuration is that the interacting sub-systems can be
  separated before measurement of their respective populations. Correlated
  current readout measures the population in the position basis.}
\label{fig:SAWSys}
\end{figure}

The simplest system to study that of two spin clusters coupled via an
Ising-like interaction,
\begin{equation}
  H=
\vec{b}_1\cdot\vec{\sigma}^1+\vec{b}_2\cdot\vec{\sigma}^2
+\frac{J}{2}\sigma_z^1\otimes\sigma_z^2
\end{equation}
where $\vec{b}_i$ are local magnetic fields and $J$ is the interaction
energy between the two clusters. We can map this system to a pair of
pseudo-spins such as double quantum dots with a excess electrons whose
localisation in one or the other dot defines the pseudo-spin vector
(Fig.~\ref{fig:DQDSys} or Fig.~\ref{fig:SAWSys}). A single electron transistor
measures the total charge excess in one dot or the other.

To implement the separation of the two sub-systems before measurement, the
device in Fig.~\ref{fig:SAWSys} could be used. A surface acoustic wave (SAW),
produced in a piezoelectric material by an inter-digital transducer driven by
a microwave generator, induces a travelling sinusoidal electric field in a
2-dimensional electron gas (2DEG) situated just below the surface of a
modulation-doped GaAs-AlGaAs semiconductor. Bias electrodes on the surface of
the device deplete the 2DEG of conduction electrons on the region of the
system.

By suitable electrode potentials at the entrance of the system, an exact
number of electrons can be transported in moving quantum dots, along quasi-1D
quantum channels (defined by surface electrodes) through the device.  Double
well potentials can be created with a specified number of electrons in each,
and by using suitable electrode geometry, made to interact via an Ising
coupling. Travelling along a sufficiently long channel, the sub-systems are
allowed to thermalise, after which the two sub-systems are rapidly separated
and read-out by simply measuring the current via Ohmic contacts.


\begin{figure}
\includegraphics[width=0.45\textwidth]{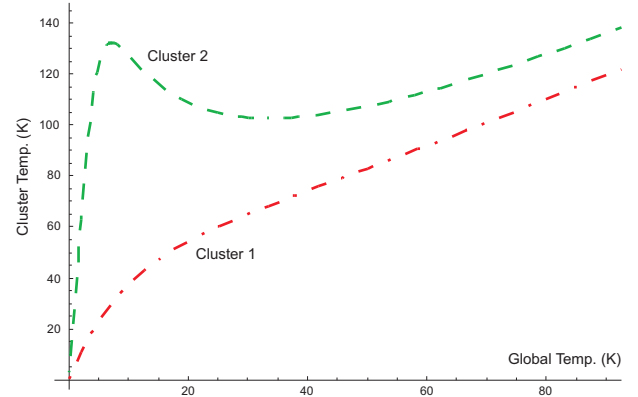}
\caption{
  A plot of the temperatures of two interacting sub-systems consisting
  of several electrons with Coulomb exchange energy.}
\label{fig:TPlot}
\end{figure}



\begin{acknowledgments}
  JO acknowledges the support of the Royal Society, and EU
  grants QAP and COSLAB. 
DKLO acknowledges the CMI programme on quantum information, Fujitsu, EU
grants TOPQIP (IST-2001-39215), RESQ (IST-2001-37559), Sidney Sussex
College, Cambridge, QIPIRC (EPSRC UK), and SUPA.
\end{acknowledgments}

\bibliographystyle{apsrev}

\end{document}